 \documentclass[twocolumn]{aastex631}

\usepackage{tabularx}
\usepackage{threeparttable}

\usepackage{rotating}
\usepackage{amsmath}
\usepackage{amssymb}
\usepackage{mathrsfs}

\begin{document}

\title[\textit{NICER} Magnetar Burst Catalog]{\textit{NICER} Magnetar Burst Catalog}

\author[0009-0002-2845-2343]{Che-Yen Chu}
\email{cychu@gapp.nthu.edu.tw}
\affiliation{Department of Physics, National Changhua University of Education, Changhua 500207, Taiwan}

\author[0000-0001-8551-2002]{Chin-Ping Hu}
\affiliation{Department of Physics, National Changhua University of Education, Changhua 500207, Taiwan}

\author[0000-0003-1244-3100]{Teruaki Enoto}
\affiliation{Department of Physics, Kyoto University, Kitashirakawa Oiwake, Sakyo, Kyoto 606-8502, Japan}
\affiliation{RIKEN Center for Advanced Photonics (RAP), 2-1 Hirosawa, Wako, Saitama 351-0198, Japan}

\author[0000-0002-7991-028X]{George A. Younes}
\affiliation{Astrophysics Science Division, NASA Goddard Space Flight Center, Greenbelt, MD 20771, USA}
\affiliation{Center for Space Sciences and Technology, University of Maryland, Baltimore County, Baltimore, MD 21250, USA}

\author[0000-0002-0254-5915]{Rachael Stewart}
\affiliation{Department of Physics, The George Washington University, Washington, DC 20052, USA}

\author{Zaven Arzoumanian}
\affiliation{Astrophysics Science Division, NASA GSFC, 8800 Greenbelt Rd., Greenbelt, MD 20771, USA}

\author[0000-0002-6089-6836]{Wynn C. G. Ho}
\affiliation{Department of Physics and Astronomy, Haverford College, 370 Lancaster Avenue, Haverford, PA 19041, USA}

\author[0000-0002-9249-0515]{Zorawar Wadiasingh}
\affiliation{Department of Astronomy, University of Maryland, College Park, Maryland 20742, USA}
\affiliation{Astrophysics Science Division, NASA Goddard Space Flight Center,Greenbelt, MD 20771, USA}
\affiliation{Center for Research and Exploration in Space Science and Technology, NASA/GSFC, Greenbelt, Maryland 20771, USA}

\author[0000-0002-0118-2649]{Andrea Sanna}
\affiliation{University of Cagliari, 09042, Monserrato, Sardinia, Italy}

\author[0000-0002-6449-106X]{Sebastien Guillot}
\affiliation{Institut de Recherche en Astrophysique et Plan\'etologie, UPS-OMP, CNRS, CNES, 9 avenue du Colonel Roche, BP 44346, Toulouse
Cedex 4, 31028, France}

\author[0000-0002-6789-2723]{Gaurava K. Jaisawal}
\affiliation{DTU Space, Technical University of Denmark, \text{\O}rsteds Plads 348, DK-2800 Lyngby, Denmark}

\author[0000-0003-4433-1365]{Matthew G. Baring}
\affiliation{Department of Physics and Astronomy - MS 108, Rice University, 6100 Main Street, Houston, Texas 77251-1892, USA}

\author[0009-0006-3567-981X]{Marlon L. Bause}
\affiliation{Max Planck Institut f\"ur Radioastronomie, Auf dem H\"ugel 69, 53121 Bonn, Germany}

\author[0000-0002-3531-9842]{Tolga G\"uver}
\affiliation{Istanbul University, Science Faculty, Department of Astronomy and Space Sciences, Beyazıt, 34119, Istanbul, Turkey}
\affiliation{Istanbul University Observatory Research and Application Center, Istanbul University 34119, Istanbul Turkey}

\author[0000-0003-1443-593X]{Chryssa Kouveliotou}
\affiliation{Department of Physics, The George Washington University, Washington, DC 20052, USA}

\author[0000-0002-3905-4853]{Alex Van Kooten}
\affiliation{Department of Physics, The George Washington University, Washington, DC 20052, USA}

\author{Keith C. Gendreau}
\affiliation{Astrophysics Science Division, NASA GSFC, 8800 Greenbelt Rd., Greenbelt, MD 20771, USA}



\begin{abstract}

In this paper, we present a comprehensive catalog of short bursts from magnetars based on eight years of \textit{NICER} observations. A total of 1130 bursts were identified from 14 sources, with the sample dominated by SGR~1935+2154, which accounts for 76\% of all detected bursts. We analyzed burst durations, spectral properties, and their correlations across multiple sources. Bursts from SGR~1935+2154 exhibit significantly longer durations, with a distribution peak at 316\,ms, compared to a peak of 23\,ms for bursts from other magnetars. Two $\mu$s-scale bursts were detected for the first time, originating from 1E~1048.1$-$5937 and CXOU~J010043.1$-$721134. Spectral analysis in the 0.5--8\,keV range using both blackbody and power-law models shows that bursts with higher fluences have harder spectra. In contrast, correlations between burst duration and spectral parameters are weak or absent. This catalog provides a valuable dataset for studying magnetar short bursts, enabling future modeling efforts and improving our understanding of the diversity and physical mechanisms of magnetar bursts.

\end{abstract}



\section{Introduction} \label{sec:intro}

Magnetars are a class of neutron stars characterized by their extremely strong magnetic fields, often on the order of $10^{14}$–$10^{15}$ G \citep{1992herm.book.....M,1992ApJ...392L...9D,1992AcA....42..145P}. In the X-ray band, unlike rotation-powered pulsars, which typically show steady emission, magnetars are known for their highly variable and transient behavior \citep[e.g.,][]{2018MNRAS.474..961C}. The first discovered magnetar burst was the giant flare from SGR~0526$-$66 in 1979, with a duration of about 10 minutes \citep{1979Natur.282..587M}. Shortly afterwards, a series of short bursts, with durations ranging from milliseconds to seconds, was detected from SGR~1900+14 \citep{1979SvAL....5..343M}. Because these repeating events are dominated by emission in the soft $\gamma$-ray band, the sources were named soft gamma repeaters (SGRs). During a magnetar outburst, which can last from several months to several years, short bursts often appear in clusters known as burst forests or burst storms, with rates that can reach hundreds of bursts per hour \citep[e.g.,][]{2020ApJ...904L..21Y,2025ApJ...989...63H}. In contrast, during the quiescent state, bursts may appear individually \citep[e.g.,][]{2002Natur.419..142G}. According to the magnetar model, these bursts are generally thought to be triggered by activity in the neutron star crust, induced by extreme magnetic stresses \citep{1984Natur.310..121L,1992ApJ...392L...9D,1995MNRAS.275..255T}.

With the discovery of more SGRs, several short-burst catalogs of individual sources have been reported, including SGR~1806$-$20, SGR~1900+14, 1E~1547.0$-$5408, SGR~0501+4516, etc. \citep{2001ApJ...558..228G,2012ApJ...749..122V,2013ApJ...778..105L}. Most of these studies focused on bursts in the hard X-ray band ($>$10\,keV), while some extended to the soft X-ray range ($<$10\,keV). The burst durations are typically log-normally distributed over more than two orders of magnitude, peaking around 200\,ms. A clear correlation between spectral hardness and brightness has been observed in these catalogs, where bursts with higher fluence tend to exhibit harder spectra, regardless of the adopted spectral model \citep{2001ApJ...558..228G,2012ApJ...749..122V,2013ApJ...778..105L}. Their X-ray spectra can generally be described equally well by models such as a double blackbody, optically thin thermal bremsstrahlung, or a Comptonized model \citep{2012ApJ...749..122V}. However, some bursts exhibit more thermal-like spectra that are adequately fitted only by a double blackbody or optically thin thermal bremsstrahlung model \citep{2008ApJ...685.1114I}.

A more comprehensive multi-source burst catalog was later compiled using data from \textit{Fermi}/Gamma-ray Burst Monitor (GBM), which included bursts from more than eight magnetar sources, with the majority originated from 1E~1547.0$-$5408 \citep{2015ApJS..218...11C}. More recently, the \textit{INTErnational Gamma-Ray Astrophysics Laboratory} (\textit{INTEGRAL})/Imager on Board the INTEGRAL Satellite (IBIS) combined over 20 years of observations to construct a catalog of 1349 magnetar bursts \citep{2026ApJ...997..272P}. The timing and spectral properties reported in these catalogs are consistent with those obtained from earlier observations. In the 2020s, several magnetars underwent new outbursts that produced numerous short bursts. Detailed analyses have been conducted for bursts from Swift~J1555.2$-$5402, SGR~1830$-$0645, and SGR~1935+2154 \citep{2021ApJ...920L...4E,2022ApJ...924..136Y,2025ApJ...989...63H}. While their general properties are similar to those of other magnetar bursts, the events from SGR~1935+2154 stand out with substantially longer durations, peaking around 1\,s, which is significantly longer than other magnetars \citep{2020ApJ...904L..21Y}.

The longer-duration short bursts were observed during the 2020 outburst of SGR~1935+2154. During this outburst, a fast radio burst (FRB)-like event FRB~200428, was detected and confirmed to be associated with a short burst from SGR~1935+2154 \citep{2020Natur.587...54C,2020Natur.587...59B,2020ApJ...898L..29M}. This discovery established SGR~1935+2154 as the only Galactic magnetar known to emit FRB-like bursts. Recently, another FRB-like event, FRB~221014A, from SGR~1935+2154 was also found to be associated with an X-ray burst from the same magnetar \citep{2026MNRAS.546ag312W}. The FRB~200428 associated X-ray short burst has a similar duration to other short bursts from this source \citep{2021NatAs...5..378L}, however, its spectrum is significantly softer \citep{2021NatAs...5..408Y}. Beyond the distinctions seen within SGR~1935+2154 itself, \citet{2013ApJ...778..105L} reported systematic differences in duration distributions between magnetars. Together, these findings highlight the diversity of magnetar burst behavior, suggesting that physical conditions and emission mechanisms vary among sources and outbursts. A systematic comparison of bursts from multiple magnetars is therefore essential to constrain the physical origin and diversity of magnetar short bursts.

In this paper, we present a systematic study of magnetar short bursts using eight years of \textit{Neutron Star Interior Composition Explorer} (\textit{NICER}) observations. We compile a catalog of bursts from multiple magnetars, including the highly active SGR~1935+2154, as well as the magnetar-like pulsars, which are high-magnetic-field rotation-powered pulsars that exhibit magnetar-like bursts \citep{2008Sci...319.1802G,2016ApJ...829L..21A,2016ApJ...829L..25G}. We perform a uniform timing and spectral analysis of these events and investigate correlations among their properties, as well as variations between different sources. This comprehensive analysis provides a uniform catalog of magnetar short bursts observed by \textit{NICER}, offering a resource for future modeling and theoretical studies.

\begin{table*}[t!]
\centering
\caption{\textit{NICER} observations used for each magnetar and magnetar-like pulsar, along with the total number of short bursts detected from each source.}
\label{tab:data}
\begin{tabular}{c|ccc}
\hline
\hline
Magnetars               & Number of observations    & Exposure (ks)  & Number of bursts  \\
\hline
CXOU J010043.1$-$721134 & 241           & 725.0     & 4    \\
4U 0142+61              & 166           & 330.0     & 4    \\
SGR 0418+5729           & 16            & 21.3      & 0    \\
SGR 0501+4516           & 198           & 498.9     & 0    \\
1E 1048.1$-$5937        & 319           & 383.2     & 1    \\
1E 1547.0$-$5408        & 60            & 87.9      & 0    \\
Swift J1555.2$-$5402    & 68            & 118.1     & 74   \\
PSR J1622$-$4950        & 22            & 46.0      & 0    \\
CXOU J164710.2$-$455216 & 111           & 166.9     & 0    \\
1RXS J170849.0$-$400910 & 113           & 84.9      & 1    \\
CXOU J171405.7$-$381031 & 1             & 3.7       & 0    \\
SGR J1745$-$2900        & 55            & 91.9      & 0    \\
SGR 1806$-$20           & 15            & 37.2      & 4    \\
XTE J1810$-$197         & 168           & 514.2     & 1    \\
Swift J1818.0$-$1607    & 74            & 181.2     & 27   \\
Swift J1822.3$-$1606    & 12            & 14.7      & 1    \\
SGR 1830$-$0645         & 78            & 245.1     & 128  \\
1E 1841$-$045           & 170           & 386.6     & 15   \\
SGR 1900+14             & 1             & 0.4       & 0    \\
SGR 1935+2154           & 415           & 1094.     & 865  \\
1E 2259+586             & 149           & 179.7     & 1    \\                    
PSR J1119$-$6127$^{*}$  & 111           & 157.0     & 0    \\
PSR J1846$-$0258$^{*}$  & 202           & 648.9     & 4    \\ 
\hline  
Total                   & 2765          & 6016.9    & 1130 \\  
\hline
\multicolumn{3}{l}{$^{*}$ Magnetar-like pulsars}
\end{tabular}
\end{table*}

\section{Data and Analysis} \label{sec:data}
\subsection{Data Selection and Reduction}
\textit{NICER} is a soft X-ray (0.2--12\,keV) telescope mounted on the International Space Station \citep{2016SPIE.9905E..1HG}. Its X-ray Timing Instrument (XTI) consists of 56 Focal Plane Modules (FPMs), though only 52 modules were in operation. While it does not offer imaging capability, \textit{NICER} provides exceptionally high time resolution ($\sim$ 300\,ns) and a large effective area, making it ideal for studying transient events. For this work, we selected all \textit{NICER} observations conducted from its launch, 2017 June, through 2025 June 30 that included a magnetar or a magnetar-like pulsar within the field of view. Among the 26 known magnetars \citep{2014ApJS..212....6O}\footnote{McGill Online Magnetar Catalog \url{http://www.physics.mcgill.ca/~pulsar/magnetar/main.html}}, \textit{NICER} has observed 21 of them, along with two additional magnetar-like pulsars. We collected 6016.9\,ks of exposure time from 2765 \textit{NICER} observations across 23 sources. The total exposure time and number of observations used for each source are summarized in Table~\ref{tab:data}.

We used \textsc{heasoft} version 6.34 \citep{2014ascl.soft08004N} to reduce the \textit{NICER} data. The standard calibration and screening were performed by the task \verb'nicerl2' with default options. The task \verb'barycorr' was then applied with the JPLEPH.430 ephemeris to calibrate the time to Barycentric Dynamical Time (TDB). The resulting event lists were separated into the standard (0.5--8\,keV) and high-energy (10--12\,keV) bands for magnetar burst searches. On 2023 May 22, \textit{NICER} sustained damage to its optical blocking film, leading to optical light leakage that degraded data quality. Observations taken after this date were processed separately. Nighttime data were reduced using the standard procedure described above. For daytime data, the \verb'nicerl2' task was executed with a screening option of \verb'threshfilter=DAY' to account for increased optical contamination.

\begin{table}[t!]
\centering
\caption{Details of \textit{NICER} observations on \textit{RXTE} blank sky regions after excluding high-background intervals.}
\label{tab:rxte}
\begin{tabular}{cccc}
\hline
\hline
Sky region & ObsID  & Exposure (ks) & Counts \\
\hline 
RXTE1   & 6012010106    & 3.1       & 1916  \\ 
RXTE2   & 1012020169    & 18.6      & 6924  \\
RXTE3   & 1012030123    & 5.3       & 2812  \\
RXTE4   & 5012040217    & 8.3       & 4336  \\
RXTE5   & 1012050131    & 5.8       & 4234  \\
RXTE6   & 4012060255    & 13.3      & 7400  \\
RXTE8   & 1012080190    & 12.2      & 5574  \\
\hline
\end{tabular}
\end{table}

\begin{figure}[t!]
\includegraphics[width=\columnwidth]{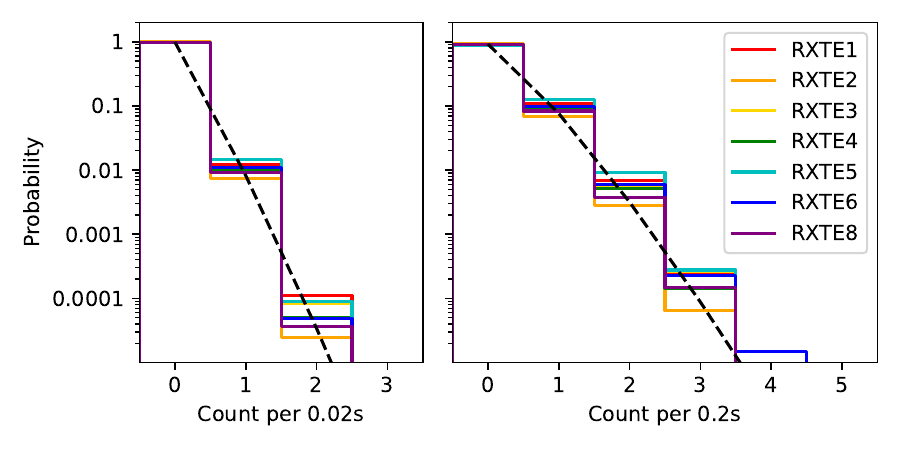}
\caption{Distributions of background counts calculated with 0.02\,s (left) and 0.2\,s (right) time bins. Different colors correspond to the RXTE blank-sky regions identified in the legend. Black dashed lines are the Poisson probability mass function of averaged background counts calculated from Table~\ref{tab:rxte}.}
\label{fig:bkg}
\end{figure}

\subsection{Burst Search} \label{sec:searching}
We applied the Bayesian block algorithm implemented in the \texttt{astropy} package, using the \verb'fitness=events' option, to both the standard and high-energy event lists to identify significant flux variations \citep{1998ApJ...504..405S,2013ApJ...764..167S}. Blocks occurring during high-particle-background intervals, as defined by the high-energy band, were excluded. In the standard band, blocks with durations shorter than 1\,s were selected as burst candidates. We further examined the FPM distribution of each candidate. Candidates with events concentrated in only one or two FPMs were identified as particle-induced events and excluded from the burst sample.

For each burst candidate, we estimated the statistical significance by computing the Poisson probability of detecting the observed number of counts within the burst block. A nearby interval of $\leq$ 5\,s was selected to represent the non-burst count rate. After accounting for the number of trials, we retained only those blocks with a Poisson probability less than $10^{-5}$ for further analysis. The number of trials was defined as the total exposure time of each observation divided by 0.2\,s, which is a typical magnetar burst duration \citep{2015ApJS..218...11C}.

\textit{NICER} has been monitoring seven X-ray blank regions defined initially by the \textit{Rossi X-ray Timing Explorer} \citep[\textit{RXTE};][]{2006ApJS..163..401J}. We selected one observation from each region and excluded intervals with high particle background. The details of the blank sky observations used are summarized in Table~\ref{tab:rxte}. The RXTE7 region is excluded because it was later found to contain a faint X-ray source \citep{2022AJ....163..130R}. In addition, some of the remaining regions contain fainter sources (up to 3\,$\times 10^{-13}$\,erg\,s$^{-1}$\,cm$^{-2}$) according to the extended ROentgen Survey with an Imaging Telescope Array (eROSITA) source catalog \citep{2024A&A...682A..34M}. We binned the cleaned event lists into 0.2\,s and 0.02\,s bins, matching the burst durations reported in previous work \citep{2015ApJS..218...11C} as well as those found in this study. We then recorded the total counts in each bin to characterize the background X-ray distribution. The cumulative distribution function reaches 100\% at 4 counts per 0.2\,s bin and 3 counts per 0.02\,s bin (Figure~\ref{fig:bkg}). To ensure robust separation between genuine bursts and background fluctuations, we required that a burst candidate contain at least 5 counts within a Bayesian block. Based on the measured background rate and the \textit{NICER} effective area, the expected detection limits for a transient event are approximately 2 and 2.5\,$\times 10^{-11}$\,erg\,cm$^{-2}$ for transient duration 0.02\,s and 0.2\,s respectively.

\subsection{Analysis}
We first plotted the cumulative counts of each selected burst, after subtracting the non-burst count rate, and fitted the resulting curve with a step function \citep{2001ApJ...558..228G}. The step height corresponds to the total number of counts in the burst. This total count was then used to compute the 5\% and 95\% levels of the cumulative distribution. The time interval between when the cumulative count reaches 5\% and 95\% is defined as the $T_{90}$ duration \citep{1993ApJ...413L.101K}. For bursts with fewer than 20 total counts, the 5\% fractional level corresponds to less than one count, making $T_{90}$ not accurately defined. In these cases, $T_{90}$ is instead defined as the duration of the corresponding Bayesian block.

The Bayesian blocks interval of each burst defines its good time interval (GTI), and the counts within this GTI were used for the spectral analysis. We first calculated the hardness ratio (HR), defined as
\begin{equation}
HR = \frac{\rm hard - \rm soft}{\rm hard + \rm soft} ,
\end{equation} 
where \textit{soft} denotes photons in the 0.5--4\,keV range and \textit{hard} denotes photons in the 4--8\,keV range. Some bursts show no photons below 2\,keV because of strong absorption along the line of sight, so we adopted 4\,keV as the boundary between the soft and hard photons, rather than the more typical 2\,keV value.

\begin{figure*}[t!]
\includegraphics[width=\textwidth]{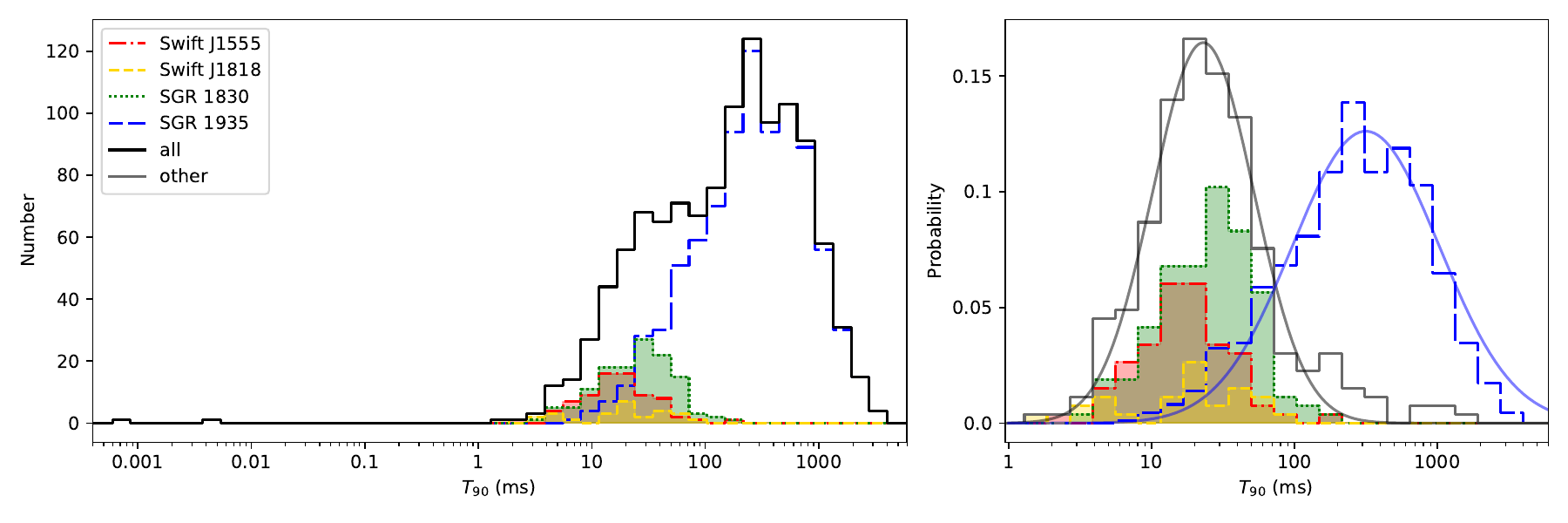}
\caption{The left panel shows the overall duration distribution of the NICER magnetar bursts, and the right panel focuses on bursts with durations longer than 1\,ms. In the right panel, the sample is divided into two groups: SGR~1935+2154 and all other magnetars, with each distribution normalised by its respective sample size so that their histogram sums to one. The red dash-dotted line represents Swift J1555.2$-$5402 (Swift~J1555), the yellow dashed line represents Swift~J1818.0$-$1607 (Swift~J1818), and the green dotted line represents SGR~1830$-$0645 (SGR~1830). The histograms of these three samples are filled for clarity. The blue long-dashed line corresponds to SGR~1935+2154 (SGR~1935), and the black solid histogram represents all bursts. The gray histogram shows all samples except SGR~1935+2154. The blue and the gray curves indicate the log-normal fits to the normalized SGR~1935+2154 and other samples, respectively.}
\label{fig:T90}
\end{figure*}

We also extracted 0.5--8\,keV burst spectra from the burst GTI with the task \verb'nicerl3-spect' and performed spectral fitting using \textsc{xspec} v12.14 \citep{2014ascl.soft08004N}. The non-burst interval used for estimating the burst significance was also adopted as the background spectrum for subtraction during fitting. The spectra were binned according to the total number of burst counts: for bursts with fewer than 20 counts, at least 1 count per bin; for 20--49 counts, each bin contained at least 5 counts; for 50--249 counts, at least 10 counts per bin; and for more than 250 counts, at least 25 counts per bin. Each burst spectrum was fitted with both a blackbody and a power-law model. In all fits, the hydrogen column density ($N_{\rm H}$) was fixed to values reported in the literature for each of the magnetars \citep{2023MNRAS.526.1287C,2021ApJ...920L...4E,2021MNRAS.504.5244B,2020ApJ...902....1H,2022ApJ...924..136Y,2023ApJ...952..120H}. For bursts with fewer than 20 total counts, the blackbody temperature ($kT$) or the power-law photon index ($\Gamma$) was also fixed at 1.2\,keV and 0.8, respectively. The \verb'cstat' in the \textsc{xspec} statistic was used for bursts with fewer than 250 counts, while the \verb'chi' statistic was adopted for bursts with 250 counts or more. Once the best-fit parameters were obtained, we calculated the absorbed flux for both models in the 0.5--10\,keV range. All uncertainties reported in this paper correspond to 1$\sigma$ confidence levels.

\begin{figure}[t!]
\includegraphics[width=\columnwidth]{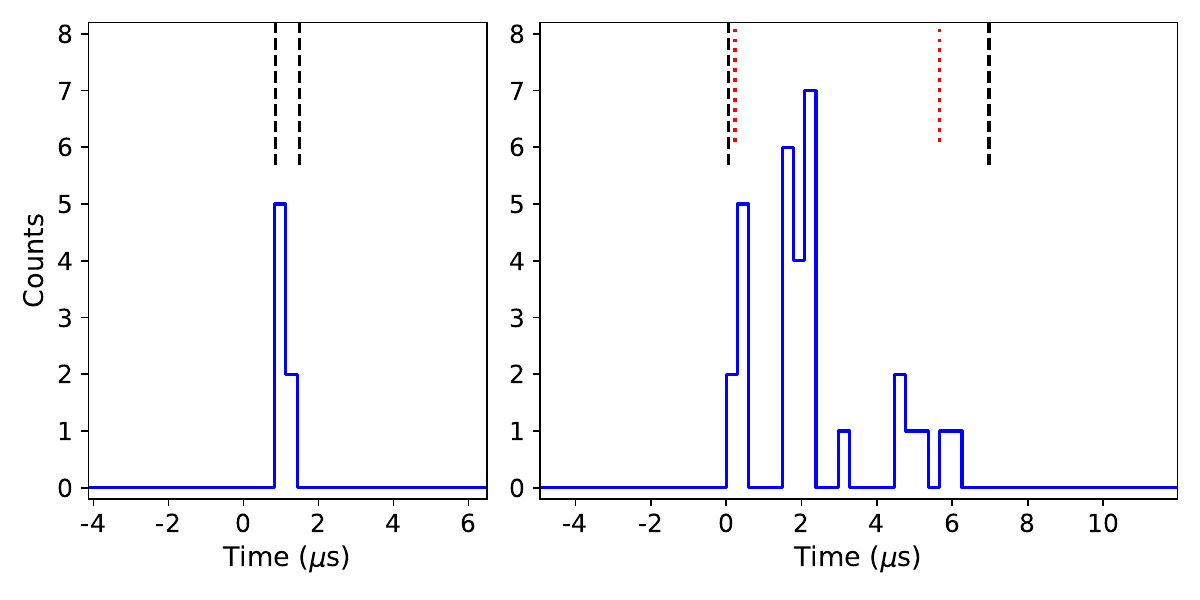}
\caption{Light curves of the $\mu$s-scale bursts with 0.3\,$\mu$s time bins. The left panel shows the burst with a $T_{90}$ of 0.6\,$\mu$s, with the zero point at 2018-11-22T11:58:14.425441 (TDB). The right panel shows the burst with a $T_{90}$ of 5\,$\mu$s, with the zero point at 2022-07-24T23:59:07.273212 (TDB). The black vertical dashed lines indicate the Bayesian block boundaries, and the red dotted lines mark the calculated $T_{90}$ intervals.}
\label{fig:mus_burst}
\end{figure}

\section{Results and Discussion}
\subsection{Overall samples}
We detected a total of 1130 bursts in our dataset, including four from the magnetar-like pulsar PSR~J1846$-$0258. The majority of events (76.5\%) originated from SGR~1935+2154, which has been the most active magnetar in recent years. \textit{NICER} captured its major outbursts in April 2020, September 2021, and October 2022 \citep{2020ApJ...904L..21Y,2021ATel14916....1G,2024Natur.626..500H,2025ApJ...989...63H}, during which some FRB-like events were also contemporaneous \citep{2020Natur.587...54C,2020Natur.587...59B,2021NatAs...5..414K,2022ATel15681....1D,2023arXiv231016932G}.

Several magnetar outbursts have also been captured by \textit{NICER} since its launch in 2017 June, including the 2020 outburst of Swift J1818.0$-$1607 \citep{2020ApJ...902....1H}, the 2020 outburst of SGR~1830$-$0645 \citep{2022ApJ...924..136Y}, the 2021 outburst of Swift~J1555.2$-$5402 \citep{2021ApJ...920L...4E}, and the 2024 outburst of 1E~1841$-$045 \citep{2025ApJ...989...89Y}. These sources accounted for 21.6\% of the bursts in our sample. \textit{NICER} also observed additional outbursts from other magnetars, such as XTE~J1810$-$197 \citep{2019ApJ...877L..30G}, and PSR~J1846$-$0258 \citep{2023ApJ...952..120H}. However, in these cases, the bursting activity was relatively weak and/or the observations did not coincide with the outburst onset, when the strongest bursting activity is typically observed.

The number of bursts detected from each source is summarized in Table~\ref{tab:data}, while details of individual bursts are provided in Appendix~\ref{sec:appendix}. Several of these bursts have also been reported in earlier studies focused on individual sources, and here we place them in the broader context of the \textit{NICER} magnetar sample.

\subsection{$T_{90}$ duration}
The distribution of $T_{90}$ durations for all bursts is shown in Figure~\ref{fig:T90}. The overall sample has a mean $T_{90}$ of 350\,ms and a standard deviation of 470\,ms. 

\subsubsection{The $\mu$s-scale bursts}
Two $\mu$s-scale bursts were identified in our sample: one from 1E~1048.1$-$5937 with a duration of 0.6\,$\mu$s, and another from CXOU~J010043.1$-$721134 with a duration of 5\,$\mu$s. We note that CXOU~J010043.1$-$721134 is located in the Small Magellanic Cloud \citep{2002ApJ...574L..29L}. We verified these $\mu$s-scale bursts are not associated with known instrumental effects. The absolute timing precision of \textit{NICER} ($\sim$ 300\,ns) is well below the measured burst durations. In each burst, the detected photons are distributed across multiple FPMs, with no single FPM registering more than one hit, which makes an instrumental or electronic origin unlikely.

To further examine whether such bursts are unique to magnetars, we applied the Bayesian Block method to all available \textit{NICER} \textit{RXTE} blank-sky region observations. Across a total exposure of 3308.9\,ks, no $\mu$s-scale blocks were identified. Given an event rate of 2 per 6016.9\,ks, the probability of observing zero events in 3308.9\,ks is 33.3\%. Although this probability is not sufficiently low to formally reject the null hypothesis, it provides additional support that the two detected $\mu$s-scale bursts are likely of astrophysical origin. We encourage future studies using \textit{NICER} data to apply the Bayesian Block method to search for similar $\mu$s-scale events.

These events represent the first detections of $\mu$s-scale X-ray bursts from magnetars. Their light curves are shown in Figure~\ref{fig:mus_burst}. Since no bursts were detected with durations between 10\,$\mu$s and 1\,ms, it remains unclear whether these events are simply the extreme tail of the main burst distribution or instead represent a distinct class of magnetar short bursts. One possible hint that $\mu$s-scale bursts differ from the main population is their extremely short duration, which may not allow sufficient time for the plasma to thermalize. Their emission is therefore expected to be non-thermal, in contrast to some normal bursts that can be described only by thermal models \citep{2008ApJ...685.1114I}. Future X-ray missions with larger effective area will be crucial to resolve this question.

\subsubsection{Distinct $T_{90}$ distributions among Magnetars}
Bursts from SGR~1935+2154 exhibit $T_{90}$ values that are significantly longer than those of other magnetars. We fitted log-normal distributions to both groups and show the normalized distributions in the right panel of Figure~\ref{fig:T90}. For SGR~1935+2154, the distribution peaks at 316\,ms, while it peaks at 23\,ms for the remaining magnetars. Thus, bursts from SGR~1935+2154 are on average about an order of magnitude longer. A Kolmogorov–Smirnov (KS) test applied to the two distributions yields a probability of $3.5 \times 10^{-106}$, confirming that the two samples are distinct. Nevertheless, a small subset of bursts from other magnetars extends to longer durations comparable to those of SGR~1935+2154, suggesting that long-duration bursts are not unique to this source and may be associated with source-dependent activity states.

\citet{2020ApJ...904L..21Y} studied bursts from SGR~1935+2154 during its 2020 outburst using \textit{NICER} data. These events are also included in our sample, accounting for roughly 40\% of the SGR~1935+2154 bursts analyzed here. However, the $T_{90}$ durations we obtain are systematically shorter than those reported in \citet{2020ApJ...904L..21Y}. A key reason is the different treatment of multi-peak bursts: we apply a 0.2\,s criterion to decide whether peaks belong to separate bursts, whereas their study used 0.5\,s. Consequently, they grouped more peaks into single multi-peak bursts, resulting in longer measured $T_{90}$ durations. A second factor is the difference in burst populations across epochs. \textit{NICER} captured a burst storm from SGR~1935+2154 in the 2020 outburst, during which the persistent emission was higher than usual. Dim or very short bursts may have been buried in this elevated background and thus missed. In contrast, the bursts detected in the 2021 and 2022 outbursts occurred during lower persistent-emission states, making weaker and shorter events easier to identify. As a result, the 2021 and 2022 bursts exhibit shorter duration distributions overall, pulling the combined distribution toward smaller $T_{90}$ values.

Regardless of methodology, the burst durations of SGR~1935+2154 differ significantly from those of other magnetars, suggesting possible intrinsic differences between this source and the broader magnetar population. SGR~1935+2154 is currently the only Galactic magnetar known to produce FRB-like bursts \citep{2020Natur.587...54C,2020Natur.587...59B}, and numerous studies have explored what physical mechanisms might make it distinct. \citet{2025ApJ...989...63H} reported that during the 2022 outburst of SGR~1935+2154, the burst properties evolved across different epochs, with a notable change after an intermediate flare. The X-ray bursts that occurred after the flare, including those around the time of the X-ray associated FRB~221014A \citep{2022ATel15681....1D,2026MNRAS.546ag312W}, exhibit properties that differ significantly from the pre-flare burst population. \citet{2021NatAs...5..408Y} also found that the X-ray counterpart of FRB~200428 exhibited a softer spectrum compared to typical bursts, potentially indicating an origin in a different region, more polar, of the magnetosphere. In this quasi-polar burst scenario, longer burst durations would be expected as field line foot points extend to larger altitudes.

These findings suggest that the differences in $T_{90}$ durations between SGR~1935+2154 and other magnetars may reflect variations in the burst physics, geometry and location across different outbursts and even between epochs within a single outburst. Our SGR~1935+2154 sample combines bursts from the 2020, 2021, and 2022 outbursts, and further disentangling these epoch-dependent effects will be necessary to understand the underlying physics. A detailed comparison of burst properties across different epochs of SGR~1935+2154 will be presented in a forthcoming paper (Chu et al., in preparation). Together, these results emphasize that SGR~1935+2154 is an exceptional case among magnetars, potentially connected to its unique ability to produce radio bursts.

In addition to bursts from other magnetars, \citet{2013ApJ...778..105L} reported 275 bursts from 1E~1547.0$-$5048 detected with \textit{Swift}/X-ray Telescope (XRT), with a $T_{90}$ distribution peaking at 207\,ms and 262 bursts from SGR~0501+4516 detected with \textit{XMM-Newton}, whose $T_{90}$ durations peak at about 100\,ms. \citet{2015ApJS..218...11C} presented a catalog of 440 bursts from at least eight magnetars observed by \textit{Fermi}/GBM. Among these, 386 bursts originate from 1E~1547.0$-$5048 and have a $T_{90}$ peak at 156\,ms, while the remaining bursts peak around 80\,ms. Overall, these duration distributions are longer than those of the regular magnetar bursts in our \textit{NICER} sample. One likely explanation is that \textit{NICER} offers superior time resolution and larger effective area than those aforementioned missions, it is more sensitive to faint or shorter duration bursts. Consequently, our catalog includes a substantial number of short bursts, leading to systematically shorter $T_{90}$ distributions compared to previous studies. 

Differences in energy coverage and background levels also play an important role. The \textit{Fermi}/GBM catalog is based on observations in the 8--200\,keV range, which differs significantly from the 0.5--8\,keV band used in this work. In addition, \textit{Fermi}/GBM has a higher background rate, which can cause the rising and decaying tails of bursts to be buried in the background. As a result, burst durations measured with GBM are often shorter than those measured with \textit{NICER}, even for the same burst, as shown by \citet{2021NatAs...5..408Y}. This effect is also evident for 1E~1547.0$-$5048, where the $T_{90}$ peak is longer in \textit{Swift}/XRT data than in \textit{Fermi}/GBM data, despite probing the same outburst \citep{2013ApJ...778..105L,2015ApJS..218...11C}. In contrast, most bursts from magnetars other than SGR~1935+2154 in our NICER catalog (e.g., Swift~J1555.2$-$5402, Swift~J1818.0$-$1607, and SGR~1830$-$0645) exhibit durations significantly shorter than those of 1E~1547.0$-$5048 detected by \textit{Fermi}/GBM. This difference suggests that the bursts captured by GBM may be primarily sourced from more energetic bursting episodes, similar to the highly active 2020 outburst of SGR~1935+2154.

\begin{table}[t!]
\centering
\caption{Log-normal fitting results to $T_{90}$ duration distributions from different magnetar groups and studies.}
\label{tab:T90_fit}
\begin{tabular}{cc|cc}
\hline
\hline
Magnetars               & Instrument               & peak  & 16--84\% quantile \\
                        &                          & (ms)  & (ms)               \\
\hline
SGR~1935+2154             & \textit{NICER}/XTI     & 316   & 95--1026          \\
Other magnetars           & \textit{NICER}/XTI     & 23    & 10--52             \\
SGR~1935+2154$^{1}$       & \textit{NICER}/XTI     & 840   & 430--1630          \\
1E~1547.0$-$5408$^{2}$    & \textit{Swift}/XRT     & 207   & 62--685            \\
1E~1547.0$-$5408$^{3}$    & \textit{Fermi}/GBM     & 156   & 62--389            \\
\hline
\multicolumn{4}{l}{References. (1) \citet{2020ApJ...904L..21Y} (2) \citet{2013ApJ...778..105L}} \\
\multicolumn{4}{l}{(3) \citet{2015ApJS..218...11C}}
\end{tabular}
\end{table}


Interestingly, the 2009 outburst of 1E~1547.0$-$5048, from which the GBM bursts originate, was also accompanied by the detection of radio bursts \citep{2021ApJ...907....7I}. Although these radio bursts were less energetic than typical extragalactic FRBs and the FRB-like bursts associated with SGR~1935+2154, their occurrence may point to the existence of different magnetar outburst modes, some of which are characterized by more energetic X-ray bursting activity and an enhanced likelihood of radio burst emission. In contrast, searches for radio burst events from several other magnetars have been conducted but yielded no detections \citep{2025A&A...700A..19G}.

A summary of the log-normal fitting results of the $T_{90}$ distributions, including the peak values and 16--84\% quantile ranges for different groups and instruments, is presented in Table~\ref{tab:T90_fit}.

\begin{figure}[t!]
\includegraphics[width=\columnwidth]{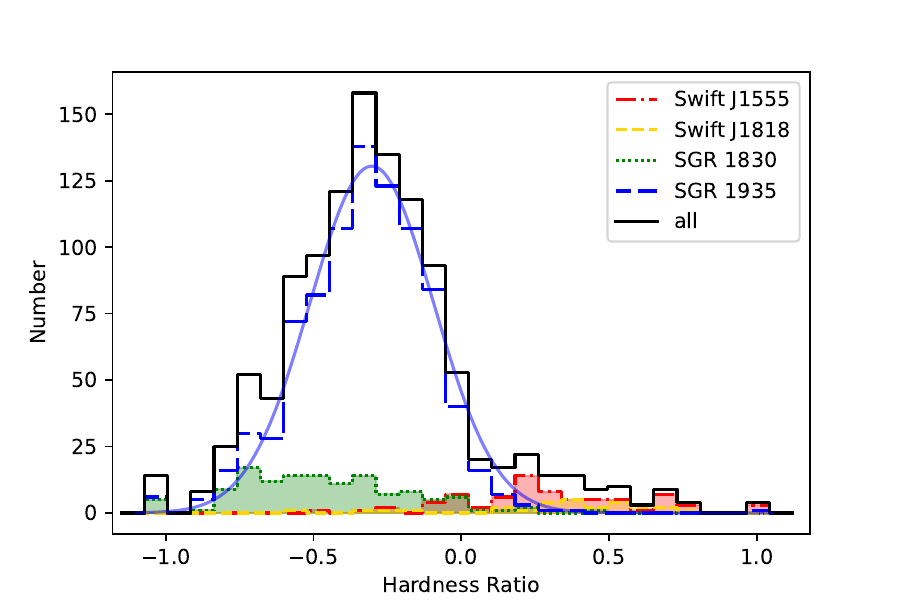}
\caption{Hardness ratio distribution of all bursts. Different histograms represent magnetar groups as defined in Figure~\ref{fig:T90} and indicated in the legend. The blue curve shows the normal distribution fit to the SGR~1935+2154 sample.}
\label{fig:HR}
\end{figure}

\subsection{Hardness ratio}
The distribution of the HR is shown in Figure~\ref{fig:HR}. The overall distribution has a mean of $-$0.29 and a standard deviation of 0.31. Most bursts exhibit relatively soft spectra, while those from Swift~J1555.2$-$5402 and Swift~J1818.0$-$1607 are noticeably harder, with HR mostly greater than zero. This trend is consistent with the higher column densities of these sources ($N_{\rm H} = 8.72$ and $11.2 \times 10^{22}$\,atoms
\,cm$^{-2}$, respectively), which preferentially absorb more soft photons compared to other magnetars.

For SGR~1935+2154, the HR distribution is well described by a normal distribution with a mean of $-$0.30 and a standard deviation of 0.21. This distribution is similar to those of other low-$N_{\rm H}$ magnetars, in contrast to the marked differences observed in their burst-duration distributions.

\subsection{Spectral model} 
The spectra of magnetar bursts in the soft X-ray band can be equally well described by either a single power-law or a single blackbody model \citep[e.g.,][]{2017ApJS..232...17K}. Therefore, each burst spectrum was fitted individually with both models to account for both possibilities.

\subsubsection{Flux} 
\begin{figure}[t!]
\includegraphics[width=\columnwidth]{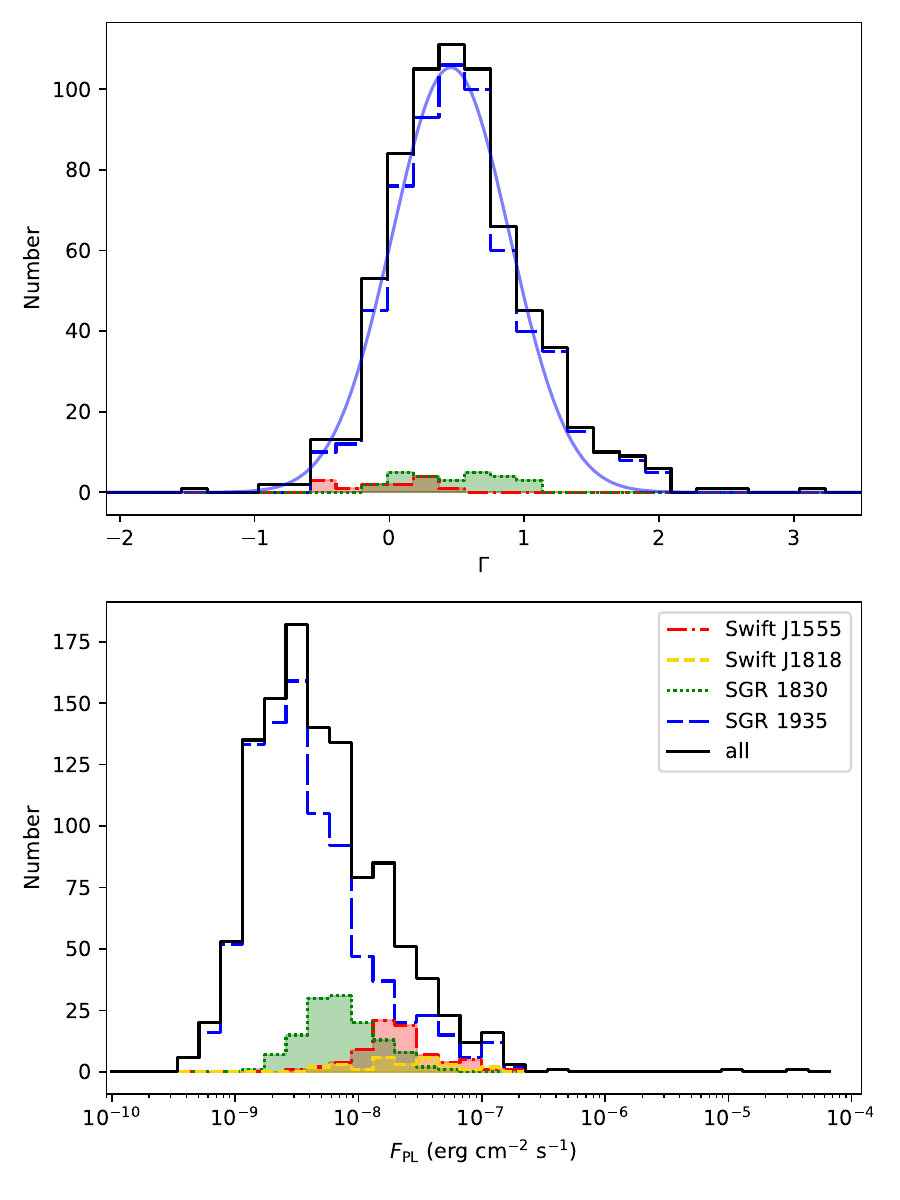}
\caption{Photon index $\Gamma$ (top) and flux $F_{\rm{PL}}$ (bottom) distributions of all bursts. Different histograms represent magnetar groups as defined in Figure~\ref{fig:T90} and indicated in the legend. For the $\Gamma$ distribution, most Swift~J1818.0$-$1607 bursts have fixed $\Gamma$ values during spectral fitting; therefore, this sample is not plotted separately on the top panel.}
\label{fig:Gm}
\end{figure}

The results of the power-law spectral fits are shown in Figure~\ref{fig:Gm}, which display the distributions of photon index ($\Gamma$) and corresponding flux ($F_{\rm{PL}}$). A normal distribution fit to SGR~1935+2154 group yields a mean $\Gamma$ of 0.45 and a standard deviation of 0.44. The photon index is slightly softer than that of \textit{Fermi}/GBM bursts \citep[$\sim$0;][]{2015ApJS..218...11C}. The difference likely reflects \textit{NICER}’s higher sensitivity to fainter bursts, which tend to exhibit softer spectra \citep{2001ApJ...558..228G,2013ApJ...778..105L}, corresponding to larger photon index (see also Section~\ref{sec:correlation}). As a result, the inclusion of additional dim bursts shifts the overall $\Gamma$ distribution toward higher values.

\begin{figure}[t!]
\includegraphics[width=\columnwidth]{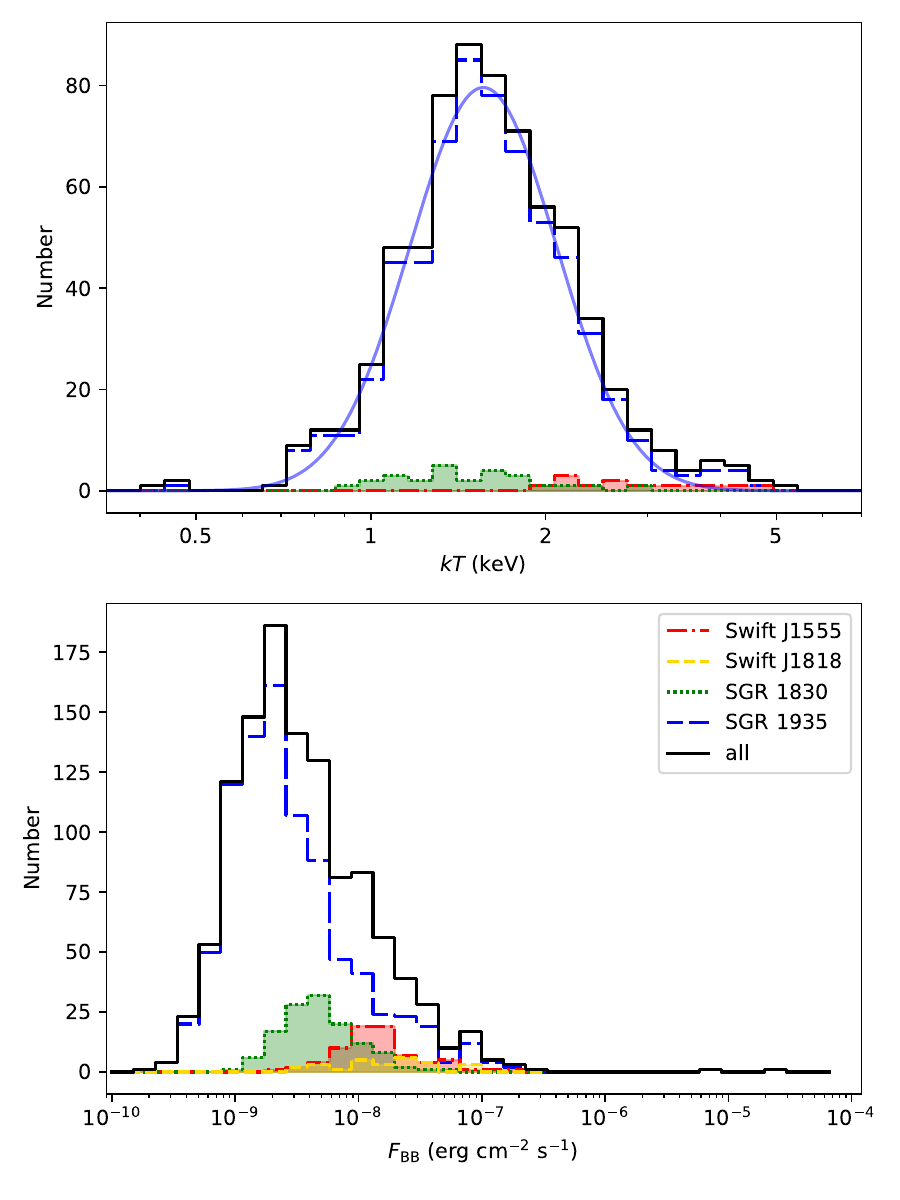}
\caption{Blackbody temperature $kT$ (top) and flux $F_{\rm{BB}}$ (bottom) distributions of all bursts. Different histograms represent magnetar groups as defined in Figure~\ref{fig:T90} and indicated in the legend. For the $kT$ distribution, most Swift~J1818.0$-$1607 bursts have fewer than 20 counts, and their $kT$ values were fixed during spectral fitting; therefore, this sample is not plotted separately on the top panel.}
\label{fig:kT}
\end{figure}

The results of the blackbody fits are shown in Figure~\ref{fig:kT}, which presents the distributions of blackbody temperature ($kT$) and flux ($F_{\rm{BB}}$). Fitting the $kT$ distribution of SGR~1935+2154 group with a log-normal function yields a peak at 1.56\,keV and a 16--84\% quantile range of 1.17--2.08\,keV. This characteristic temperature is slightly lower than that obtained from \textit{Swift}/XRT bursts of 1E~1547.0$-$5408 \citep[$\sim$2.5\,keV;][]{2013ApJ...778..105L} and considerably lower than that of \textit{Fermi}/GBM bursts \citep[$\sim$4.5\,keV;][]{2015ApJS..218...11C}. Similar selection effects likely explain the lower characteristic temperature compared to \textit{Swift}/XRT and \textit{Fermi}/GBM results. Because \textit{NICER} is more sensitive to faint, softer bursts, the inclusion of these events shifts the overall $kT$ distribution toward lower values. The higher temperatures inferred from \textit{Fermi}/GBM may also arise because those fits correspond to the lower-temperature component of a two-blackbody model in the 8–200\,keV range, where the cooler component is poorly constrained and thus biased upward. Alternatively, the broadband spectra may be better represented by three blackbody components rather than two. In any case, the burst temperatures are significantly higher than the typical magnetar temperature in quiescence of $\sim$0.45\,keV  \citep{2023MNRAS.526.1287C}, and even during the onset of outburst of $\sim$1.2\,keV \citep[e.g.][]{2020ApJ...902....1H,2021ApJ...920L...4E}. The fluxes are generally distributed between $10^{-10}$ and $3 \times 10^{-7}$\,erg\,s$^{-1}$\,cm$^{-2}$, with the two $\mu$s-scale bursts showing much higher fluxes of $\sim10^{-5}$\,erg\,s$^{-1}$\,cm$^{-2}$.

The flux distribution obtained from the power-law model is broadly consistent with that derived from the blackbody model, indicating that the inferred burst fluxes are not strongly dependent on the chosen spectral model.

\begin{figure}[t!]
\includegraphics[width=\columnwidth]{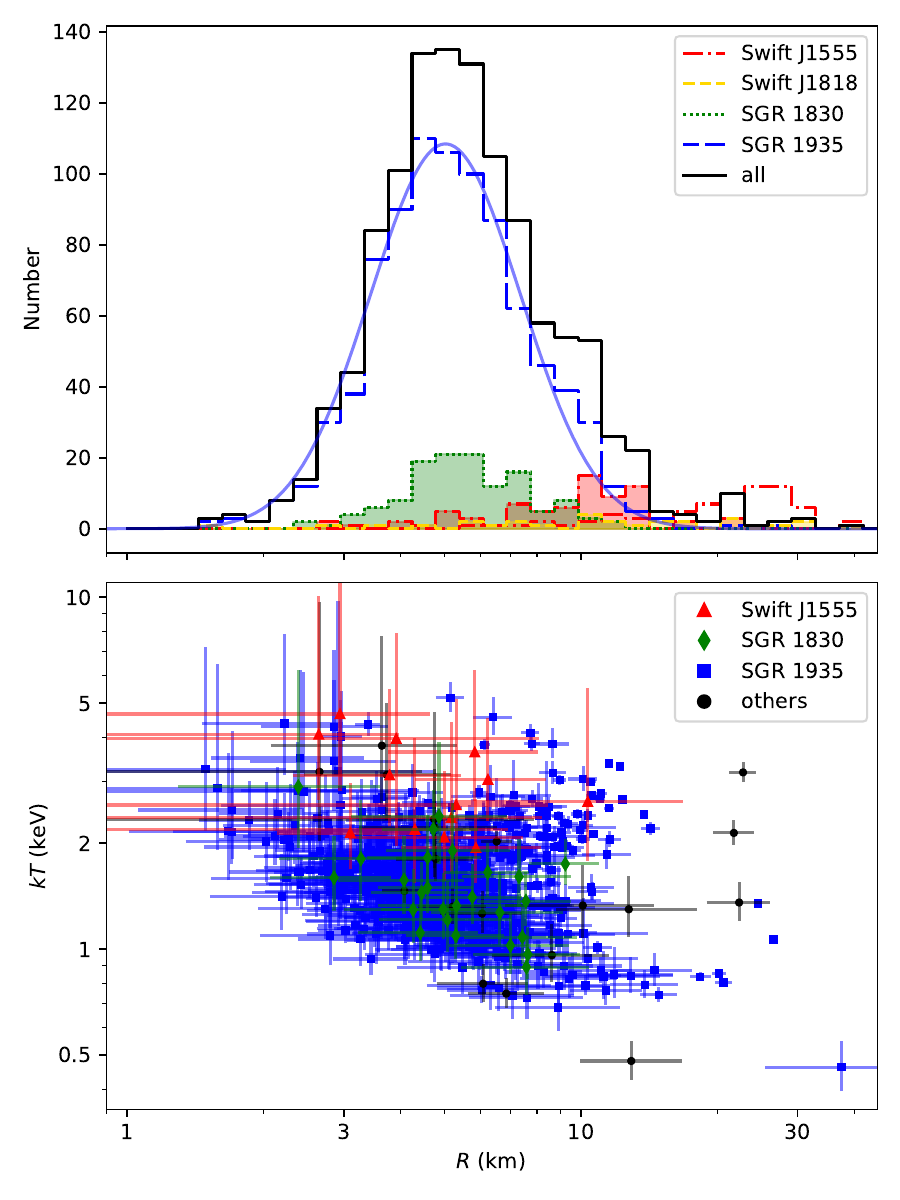}
\caption{Distribution of blackbody emission radius $R$ (top) and emission radius $R$ versus temperature $kT$ (bottom). Different histograms represent the magnetar groups defined in Figure~\ref{fig:T90}, as indicated in the legend. The unshaded red histogram shows Swift~J1555.2$-$5402 assuming a distance of 10\,kpc, while the shaded red histogram assumes a distance of 4.5\,kpc. The blue curve shows the log-normal fit to the $R$ distribution of the SGR~1935+2154 sample. In the bottom panel, red triangles represent Swift~J1555.2$-$5402, green diamonds represent SGR~1830$-$0645, blue squares represent SGR~1935+2154, and black circles represent the remaining magnetars.}
\label{fig:R}
\end{figure}

We also calculate the emission radius ($R$) from the normalization of the blackbody flux. The distribution of $R$ is shown in the top panel of Figure~\ref{fig:R}. The distances used to calculate $R$ are adopted from \citet{2014ApJS..212....6O,2023MNRAS.526.1287C} and the references therein, as well as \citet{2021ApJ...920L...4E,2022ApJ...924..136Y}.  Fitting the $R$ distribution of SGR~1935+2154 group with a log-normal function yields a peak at 5.0\,km and a 16--84\% quantile range of 3.4--7.4\,km. The size of the thermal emitting region is mostly consistent with previous work \citep{2020ApJ...904L..21Y}. Most bursts show $R$ values larger than the typical emitting size of surface thermal emission from neutron stars, about 3\,km \citep[e.g.,][]{2023MNRAS.520.4068C,2023MNRAS.526.1287C}. This suggests that the burst emission likely originates from a fireball at higher altitude rather than directly from the neutron star surface.

The $R$ distribution of Swift~J1555.2$-$5402 has a mean of 22\,km when calculated assuming a distance of 10\,kpc at the Scutum–Centaurus Arm \citep{2021ApJ...920L...4E}, which is significantly larger than that of other magnetars. We consider that this discrepancy mainly arises from the distance estimate. The placement of Swift~J1555.2$-$5402 at the Scutum–Centaurus Arm was motivated by the fact that another magnetar in the same direction, 1E~1547.0$-$5408, is estimated to lie at 4.5\,kpc near the Perseus Arm \citep{2010ApJ...710..227T}. Since Swift~J1555.2$-$5402 has an $N_{\rm H}$ value roughly twice that of 1E~1547.0$-$5408, it was suggested to reside in a more distant spiral arm \citep{2021ApJ...920L...4E}. If Swift~J1555.2$-$5402 is instead placed at a distance of 4.5\,kpc, the inferred emission radius becomes more consistent with those of other magnetars. This suggests that Swift~J1555.2$-$5402 may lie at a similar distance to 1E~1547.0$-$5408 but in a denser local environment. Even with this assumption, the $R$ distribution of Swift~J1555.2$-$5402, with a mean of about 10\,km, remains somewhat larger than that of other magnetars, possibly indicating a different emission geometry that produces a larger fireball area. The $R$ distribution assuming a distance of 10\,kpc is shown as the unshaded red histogram in Figure~\ref{fig:R}, while the distribution for 4.5\,kpc is shown as the shaded red histogram. The $R$ values of Swift~J1555.2$-$5402 reported in the Appendix tables are calculated assuming a distance of 4.5\,kpc.

The bottom panel of Figure~\ref{fig:R} shows the relation between $R$ and $kT$. An overall anti-correlation is observed, as expected for blackbody emission. We note that the $\mu$s-scale bursts are not included in this figure because their inferred emission radii are unphysically large (1000\,km and 36000\,km), which further supports a non-thermal origin for these events.

\begin{figure}[t!]
\includegraphics[width=\columnwidth]{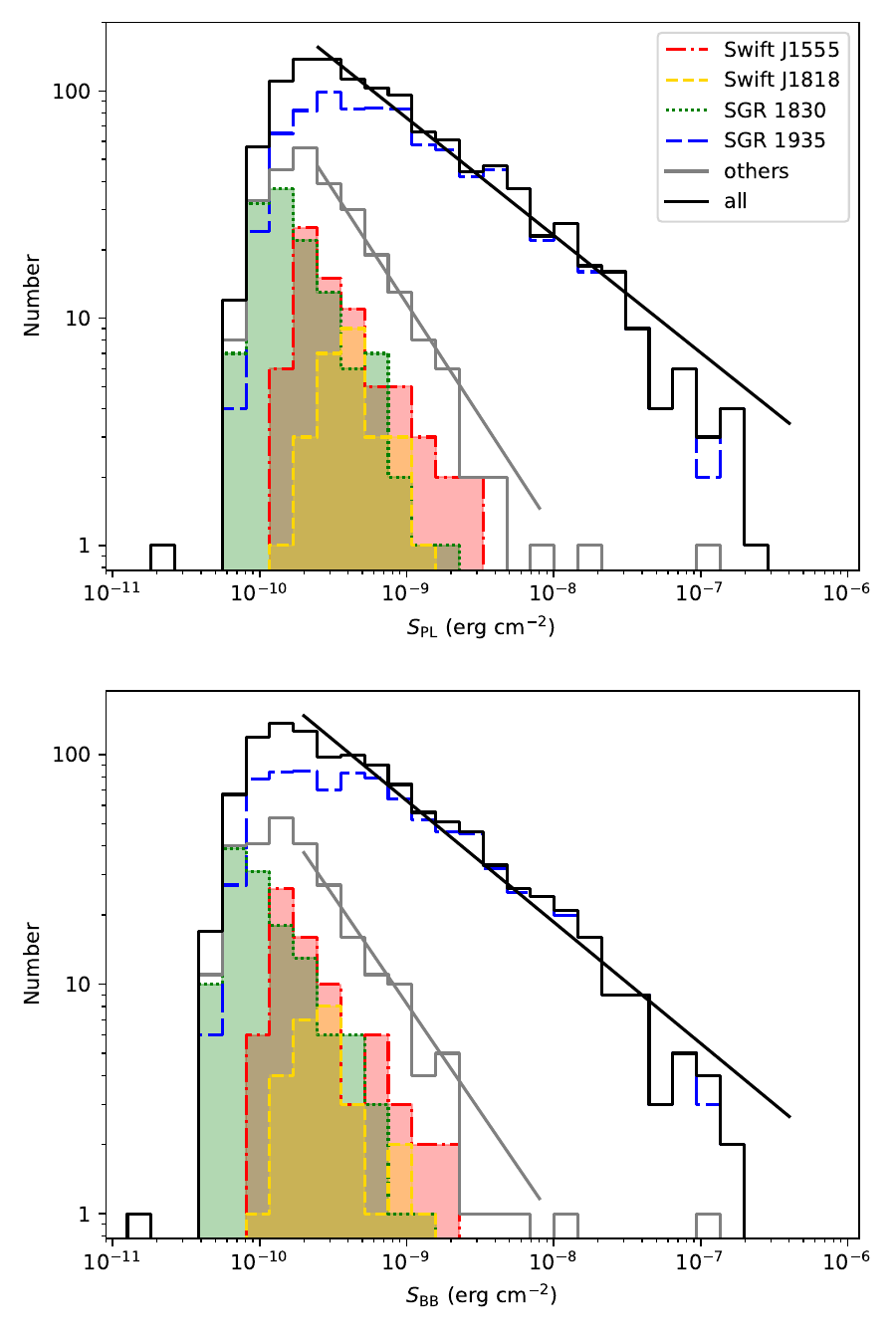}
\caption{Fluence distributions derived from the power-law model $S_{\rm PL}$ (top) and the blackbody model $S_{\rm BB}$ (bottom) for all bursts. Different histograms represent magnetar groups as defined in Figure~\ref{fig:T90} and indicated in the legend. The black lines show the best-fit power-law functions for the full burst sample, with indices of $-0.52 \pm 0.03$ for the power-law model  and $-0.53 \pm 0.04$ for the blackbody model. The gray lines show the best-fit power-law functions for the sample excluding SGR~1935+2154, with indices of $-1.00 \pm 0.06$ and $-0.94 \pm 0.06$ for the two spectral models.}
\label{fig:fluence}
\end{figure}

\begin{figure*}[t!]
\includegraphics[width=1.8\columnwidth]{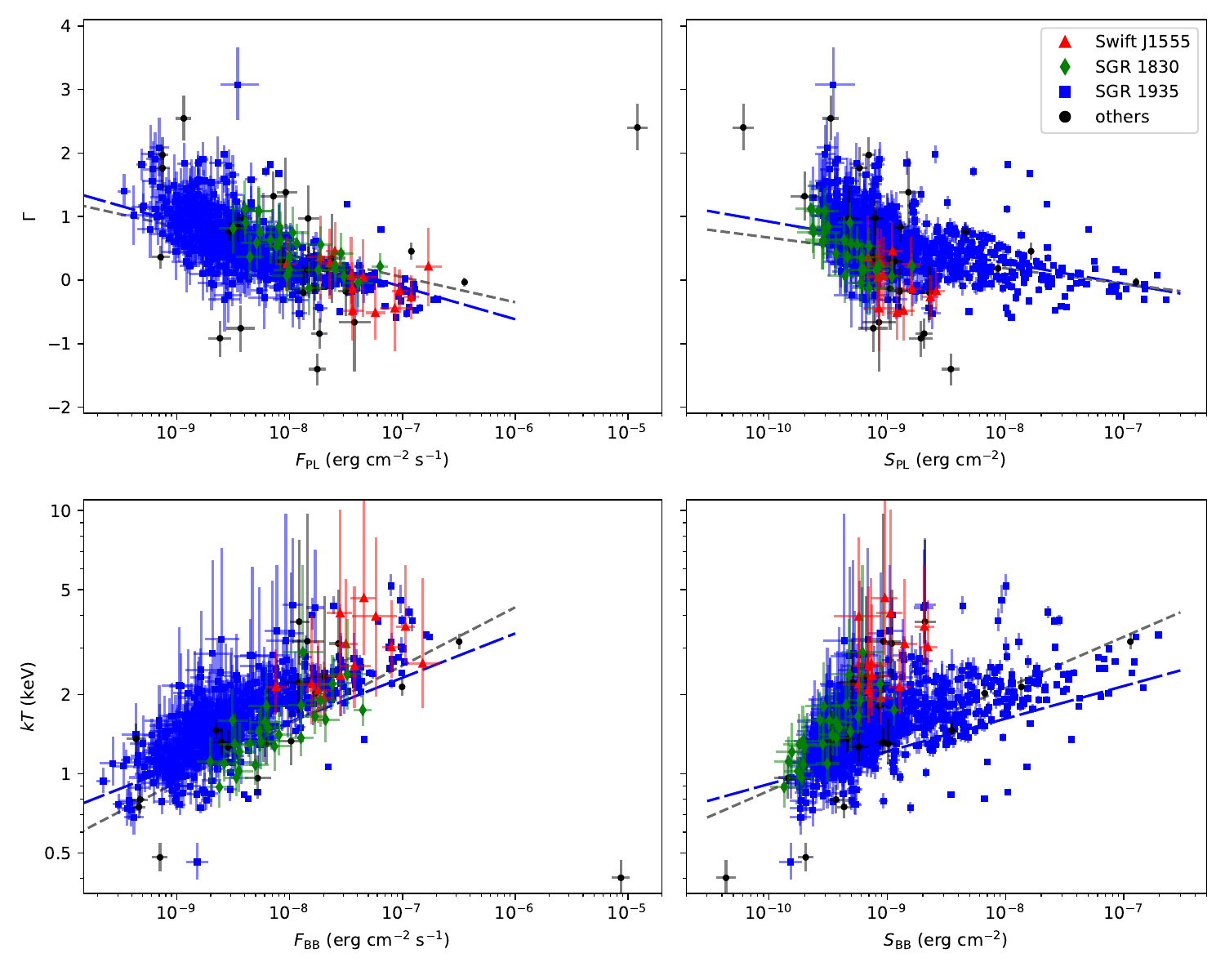}
\centering
\caption{Distribution of spectral parameters. Different data points represent magnetar groups as defined in Figure~\ref{fig:R} and indicated in the legend. The blue long-dashed lines and the gray dashed lines show the best-fit power-law relations for SGR~1935+2154 and for all other magnetars, respectively.}
\label{fig:spec}
\end{figure*}

\subsubsection{Fluence} 
The fluence ($S_{\rm{PL}}$ and $S_{\rm{BB}}$) distributions, calculated from the fluxes and $T_{90}$ durations, for both the power-law and blackbody models are shown in Figure~\ref{fig:fluence} as log($S$)–log($N$) plots. For fluences above $2 \times 10^{-10}$\,erg\,cm$^{-2}$, the distributions are well described by power-laws, yielding indices of $-$0.52 $\pm$ 0.03 for the power-law model and $-$0.53 $\pm$ 0.04 for the blackbody model. Because the full sample is dominated by SGR~1935+2154, we also fit the distributions after excluding its bursts. This yields significantly steeper slopes of $-1.00 \pm 0.06$ for the power-law model and $-0.94 \pm 0.06$ for the blackbody. These results demonstrate that SGR~1935+2154 exhibits a distinct fluence distribution compared to other magnetars. Its longer burst durations lead to a larger number of high-fluence events, which strongly influences the overall distribution. This further supports the idea that SGR~1935+2154 underwent a different outburst mode during its active episodes in 2020–2022 compared with other magnetars.

From the turnover in these distributions, we estimate that \textit{NICER} achieves nearly 100\% detection efficiency above $2 \times 10^{-10}$\,erg\,cm$^{-2}$, with a detection limit of approximately $4 \times 10^{-11}$\,erg\,cm$^{-2}$. For the duration distribution of our burst sample, this detection limit is slightly higher than the expected value of $2.5 \times 10^{-11}$\,erg\,cm$^{-2}$ estimated in Section~\ref{sec:searching}. The reason is that most of the bursts identified in this work occurred during a magnetar X-ray outburst, where the persistent emission is higher. As a result, the background level is higher, and some low-fluence bursts are buried in the background.

\begin{figure}[t!]
\includegraphics[width=\columnwidth]{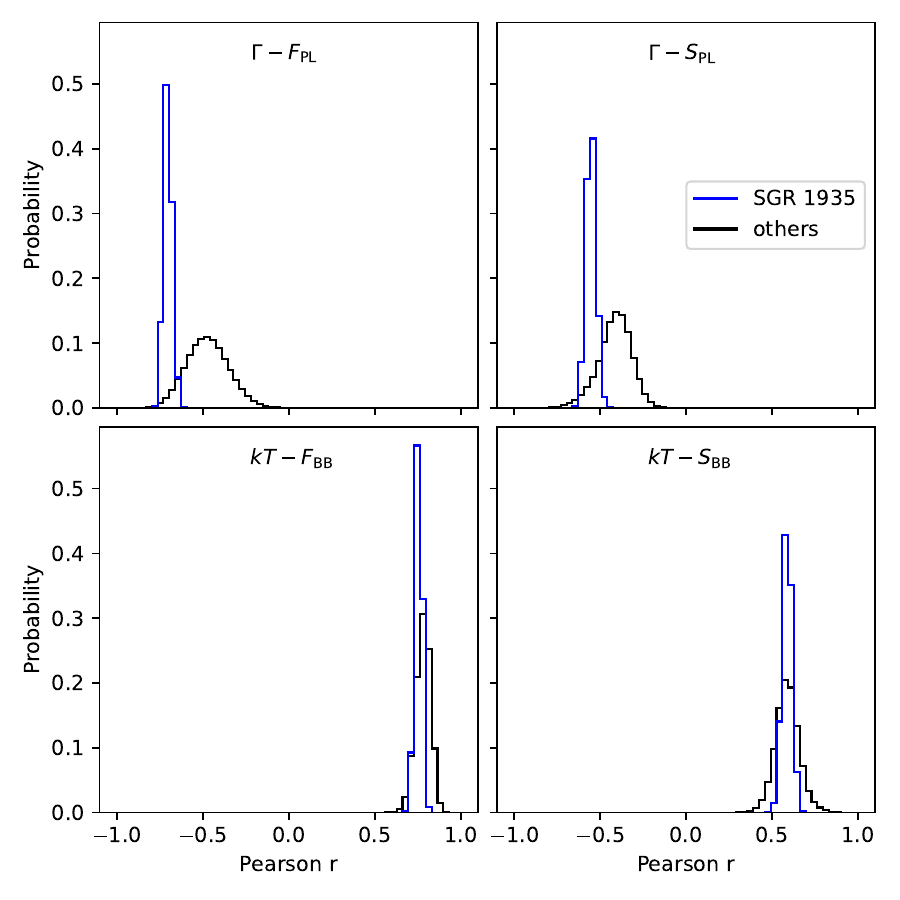}
\centering
\caption{Distributions of Pearson correlation coefficients between spectral parameters obtained using the bootstrap method. The parameter combinations are indicated at the top of each panel. The blue and the black histograms represent the SGR~1935+2154 sample and all other magnetar samples, respectively.}
\label{fig:corr_spec}
\end{figure}

\begin{figure*}[t!]
\includegraphics[width=\textwidth]{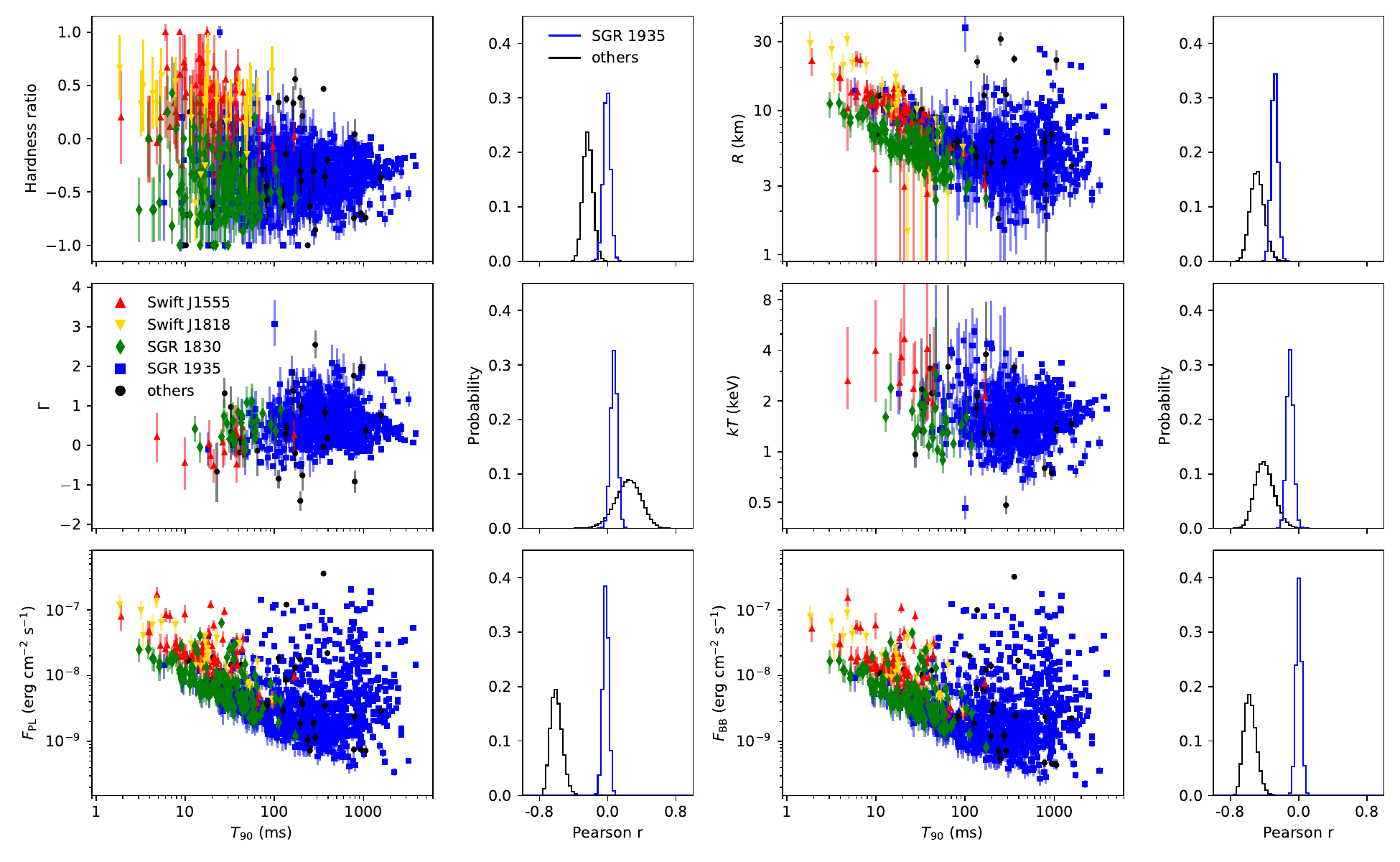}
\caption{Distributions of $T_{90}$ versus spectral parameters for all bursts (scatter plots) and the corresponding distributions of Pearson correlation coefficients obtained from bootstrap resampling (histograms). Data points in the scatter plots represent magnetar groups as defined in Figure~\ref{fig:spec}, with the yellow triangles indicating Swift~J1818.0$-$1607. The blue and the black histograms correspond to the SGR~1935+2154 sample and all other magnetars, respectively. The two $\mu$s-scale bursts are excluded from this correlation analysis.}
\label{fig:corr_time}
\end{figure*}

\subsection{Correlation}  \label{sec:correlation}
\subsubsection{Spectral vs spectral}
The left panels of Figure~\ref{fig:spec} show the distributions of $\Gamma$ and $kT$ versus their corresponding fluxes. For low-count bursts, where $\Gamma$ or $kT$ were fixed during fitting, these events are excluded from the figure and from the correlation analysis. Overall, bursts with higher fluxes tend to exhibit harder spectra, characterized by higher temperatures or smaller photon indices, consistent with previous studies \citep[see, e.g.,][]{2001ApJ...558..228G,2013ApJ...778..105L}. To quantify this relation, we calculated Pearson correlation coefficients (Pearson r) for bursts from SGR~1935+2154 and from other magnetars separately. The clear outlier, the 5\,$\mu$s burst from CXOU~J010043.1$-$721134, was excluded from this analysis. 

For SGR~1935+2154, the coefficients are $-$0.702 for $\Gamma$ versus $F_{\rm PL}$ and 0.754 for $kT$ versus $F_{\rm BB}$. For the other magnetar bursts, the corresponding values are $-$0.556 and 0.807, respectively. Strong correlations are found in all cases, except for the weaker $\Gamma$-$F_{\rm PL}$ relation in the other magnetar sample. We fitted power-law functions to these highly correlated relations, obtaining indices of 0.18 $\pm$ 0.01 and  $-$0.42 $\pm$ 0.08 for SGR~1935+2154 and the other magnetars, respectively, for the $\Gamma$-$F_{\rm PL}$ relation. For the $kT$-$F_{\rm BB}$ relation, the indices are $-$0.51 $\pm$ 0.03 for SGR~1935+2154 and 0.22 $\pm$ 0.02 for the remaining magnetars. In addition to the $\mu$s burst, two other events exhibit notably soft spectra ($kT < 0.5$\,keV or $\Gamma > 2.5$): one from SGR~1935+2154 and one from 4U 0142+61.

The right panels of Figure~\ref{fig:spec} show the distributions of $\Gamma$ and $kT$ versus $S_{\rm PL}$ and $S_{\rm BB}$, respectively. Similar to the flux relations, bursts with higher fluences generally exhibit harder spectra. For SGR~1935+2154, the Pearson correlation coefficients are $-$0.554 for $\Gamma$ versus $S_{\rm PL}$ and 0.588 for $kT$ versus $S_{\rm BB}$, while for the other magnetar bursts, they are $-$0.411 and 0.574, respectively. The correlations weaken when moving from flux to fluence, as the scatter increases at higher fluences, suggesting the presence of possible subgroups. However, after incorporating the duration parameter, the 5\,$\mu$s burst aligns with the overall trend, remaining consistent with the main burst population.

We also examined the fluence relations using power-law fits. For the $\Gamma$-$S_{\rm PL}$ relation, the derived indices are $-$0.32 $\pm$ 0.03 for SGR~1935+2154 and $-$0.28 $\pm$ 0.07 for the other magnetars. In the $kT$-$S_{\rm BB}$ relation, the indices are 0.13 $\pm$ 0.01 for SGR~1935+2154 and 0.19 $\pm$ 0.02 for the remaining sources. Because these fits are driven primarily by the high-fluence end of the distribution, the results appear weaker and less well defined than those seen in the flux relations.

To further investigate possible subgroups within our sample, we performed a bootstrap analysis of the spectral correlations. For each selected burst sample, we generated a resampled dataset by randomly selecting bursts from the original sample until the resampled dataset matched the original in size. Each burst could be selected multiple times. We then recalculated the Pearson correlation coefficient for each resampled dataset. This process was repeated one million times, and the resulting distributions of correlation coefficients are shown in Figure~\ref{fig:corr_spec}.

The mean correlation coefficients derived from the bootstrap analysis are consistent with those obtained from the original samples across all parameter combinations. However, the coefficient distributions for the other magnetar bursts are noticeably broader than those for the SGR~1935+2154 sample, particularly for the power-law model. This wider spread likely reflects the inclusion of bursts from multiple magnetars, whose spectral parameters may be influenced by instrumental limitations and varying absorption along different lines of sight. The bootstrap results show no additional outlying behavior, and no secondary clusters or anomalous distributions appear across the one million trials. This suggests that all bursts follow a similar trend, except for the 5\,$\mu$s burst in the flux relation.

\subsubsection{Timing vs spectral}
The correlations between burst duration and spectral parameters are shown in Figure~\ref{fig:corr_time}, along with their bootstrap distributions. As with the spectral–spectral relations, bursts were divided into two groups: SGR~1935+2154 and all other magnetars. The two $\mu$s-scale bursts were excluded from the analysis. For SGR~1935+2154, all correlations between $T_{90}$ and the spectral parameters are statistically insignificant, with Pearson coefficients consistent with zero. This is consistent with earlier studies of magnetar bursts, which also found no clear correlation between timing and spectral properties \citep{2012ApJ...749..122V}. 

In contrast, for the other magnetars, both $F_{\rm PL}$ and $F_{\rm BB}$ show an apparent anti-correlation with $T_{90}$. This trend is not intrinsic but instead results from instrumental limitations. Bursts with lower fluence are more likely to fall below the detection threshold, truncating the intrinsic burst population. The observed anti-correlation is therefore a selection effect. Similarly, the weak correlations and anti-correlations between $T_{90}$ and hardness ratio, $R$, $kT$, and $\Gamma$, can be explained by the same detection bias. Shorter bursts tend to have lower fluxes, and as shown in Figure~\ref{fig:corr_spec}, fainter bursts generally display softer spectra. This leads to the weak correlations between duration and spectral parameters observed here.

\section{Summary}
In this study, we constructed a comprehensive catalog of magnetar short bursts spanning eight years of \textit{NICER} observations. We identified a total of 1130 bursts and most of these bursts originated from SGR~1935+2154, which contributed 865 events. We present the distributions of timing and spectral parameters for all detected bursts and investigate their correlations. The bursts from SGR~1935+2154 generally show longer $T_{90}$ durations than those from other magnetars. We also report the first detection of $\mu$s-scale bursts in magnetars, one from 1E~1048.1$-$5937 and another from CXOU~J010043.1$-$721134.

Bursts with higher fluxes tend to exhibit harder spectra, indicating higher blackbody temperatures in the blackbody model or lower photon indices in the power-law model. The bootstrap resampling analysis confirms that these spectral correlations are statistically robust, while correlations between burst duration and spectral parameters remain weak or absent. The weaker correlations found among other magnetars likely reflect differences in source properties and instrumental effects.

This catalog provides a uniform, statistically rich sample that will serve as a foundation for future investigations of magnetar burst energetics, emission mechanisms, and magnetic field dynamics.


\begin{acknowledgments}
This research has made use of data and software provided by the High Energy Astrophysics Science Archive Research Center (HEASARC), which is a service of the Astrophysics Science Division at NASA/GSFC. C.-Y.C.
acknowledges support from the National Science and Technology Council (NSTC) in Taiwan through grant 113-2811-M-018-003-MY2 and 112-2112-M-018-004-MY3. C.-P.H. acknowledges support from the NSTC in Taiwan through grant 112-2112-M-018-004-MY3. S.G. acknowledges the support of the CNES. W.C.G.H. acknowledges support through grant 80NSSC23K0078 from NASA.
\end{acknowledgments}

%

\vspace{5mm}
\facilities{\textit{NICER}}


\software{Heasoft \citep{2014ascl.soft08004N}}




\appendix
\section{The Catalog}\label{sec:appendix}

\renewcommand{\thetable}{A\arabic{table}}
\setcounter{table}{0}

The ppendix table at the end of this paper presents example bursts identified in this study. A complete burst catalog can be found in the online version.

\bibliography{magnetar}{}
\bibliographystyle{aasjournal}

\begin{sidewaystable}
\centering
\caption{Magnetar burst catalog}
\hspace*{-2.5cm}
\begin{tabular}{c|cccccccccc}
\hline
\hline
Magnetar    & Burst      & ObsID & 	 	TDB start 	 		& $T_{90}$ 	& HR 	 		 	& $kT^{a}$  					 		& $R^{a}$ 	 		 	& $log F_{\rm BB}$ 	 						& $\Gamma$ 						& $log F_{\rm PL}$ 						 \\            &       &       & 	 	        	 		& (ms)   	&   	 		 	& (keV)					 		& (km)					 		& (erg cm$^{-2}$ s$^{-1}$)				&         						& (erg cm$^{-2}$ s$^{-1}$)			 \\
\hline
CXOU J010043.1$-$721134 & 1 	& 5505010502 	& 2022-07-24T23:59:07.273 	& 0.005 	& $-0.87\pm0.09$ 	& $0.40\substack{+0.07\\-0.05}$ 	& $35828.0^{b}$ 	& $-5.06\substack{+0.08\\-0.08}$ 	& $2.40\substack{+0.37\\-0.36}$ 	& $-4.92\substack{+0.09\\-0.09}$ \\ 
& ...	& ... 	& ... 	& ... 	& ... 	& ...  & ... 	& ... & ... & ... \\ 
& 4 	& 5505011704 	& 2023-07-31T20:21:10.490 	& 1055.592 	& $-0.74\pm0.07$ 	& $1.36\substack{+0.19\\-0.15}$ 	& $22.3\substack{+3.8\\-3.4}$ 	& $-9.36\substack{+0.09\\-0.08}$ 	& $0.36\substack{+0.18\\-0.18}$ 	& $-9.14\substack{+0.08\\-0.08}$ \\ 
\hline
4U 0142+61 & 1 	& 0020040102 	& 2017-07-15T14:15:37.947 	& 27.684 	& $-0.56\pm0.16$ 	& $0.96\substack{+0.22\\-0.16}$ 	& $8.6\substack{+2.9\\-2.3}$ 	& $-8.28\substack{+0.12\\-0.11}$ 	& $1.32\substack{+0.39\\-0.38}$ 	& $-8.14\substack{+0.12\\-0.11}$ \\ 
& ...	& ... 	& ... 	& ... 	& ... 	& ...  & ... 	& ... & ... & ... \\  
& 4 	& 6020040102 	& 2023-09-26T11:56:38.889 	& 287.419 	& $-0.86\pm0.08$ 	& $0.48\substack{+0.07\\-0.06}$ 	& $12.9\substack{+3.8\\-2.9}$ 	& $-9.15\substack{+0.07\\-0.07}$ 	& $2.54\substack{+0.36\\-0.34}$ 	& $-8.94\substack{+0.07\\-0.07}$ \\ 
\hline
1E 1048.1$-$5937 & 1 	& 1020240141 	& 2018-11-22T11:58:14.425 	& 0.001 	& $-0.71\pm0.26$ 	& $1.20$ 	& $1026.0^{b}$ 	& $-4.55\substack{+0.15\\-0.17}$ 	& $0.80$ 	& $-4.38\substack{+0.15\\-0.17}$ \\ 
\hline
Swift J1555.2$-$5402 & 1 	& 4202190101 	& 2021-06-03T13:58:12.407 	& 15.382 	& $-0.27\pm0.29$ 	& $1.20$ 	& $11.5\substack{+1.8\\-1.7}$ 	& $-7.85\substack{+0.12\\-0.14}$ 	& $0.80$ 	& $-7.67\substack{+0.12\\-0.14}$ \\ 
& ...	& ... 	& ... 	& ... 	& ... 	& ...  & ... 	& ... & ... & ... \\ 
& 74 	& 4560013017 	& 2021-10-23T06:18:22.667 	& 22.435 	& $-0.33\pm0.31$ 	& $1.20$ 	& $9.6\substack{+1.6\\-1.6}$ 	& $-8.00\substack{+0.14\\-0.15}$ 	& $0.80$ 	& $-7.82\substack{+0.14\\-0.15}$ \\ 
\hline
1RXS J170849.0$-$400910 & 1 	& 3622050401 	& 2020-04-03T18:13:10.886 	& 111.437 	& $0.34\pm0.09$ 	& $-$ 	& $-$ 	& $-7.70\substack{+0.04\\-0.06}$ 	& $-0.84\substack{+0.23\\-0.24}$ 	& $-7.74\substack{+0.07\\-0.07}$ \\ 
\hline
SGR 1806$-$20 & 1 	& 5020410102 	& 2023-02-25T00:23:44.944 	& 136.532 	& $0.37\pm0.04$ 	& $2.14\substack{+0.19\\-0.16}$ 	& $21.7\substack{+2.3\\-2.2}$ 	& $-7.00\substack{+0.03\\-0.02}$ 	& $0.45\substack{+0.14\\-0.14}$ 	& $-6.92\substack{+0.02\\-0.02}$ \\ 
& ...	& ... 	& ... 	& ... 	& ... 	& ...  & ... 	& ... & ... & ... \\ 
& 4 	& 5020410103 	& 2023-02-28T21:10:53.593 	& 355.533 	& $0.47\pm0.02$ 	& $3.18\substack{+0.23\\-0.20}$ 	& $22.8\substack{+1.6\\-1.5}$ 	& $-6.50\substack{+0.01\\-0.01}$ 	& $-0.03\substack{+0.07\\-0.07}$ 	& $-6.45\substack{+0.01\\-0.01}$ \\ 
\hline
XTE J1810$-$197 & 1 	& 2020420105 	& 2019-03-31T22:57:55.917 	& 195.301 	& $0.38\pm0.10$ 	& $-$ 	& $-$ 	& $-7.85\substack{+0.04\\-0.05}$ 	& $-1.40\substack{+0.25\\-0.26}$ 	& $-7.75\substack{+0.07\\-0.08}$ \\ 
\hline
Swift J1818.0$-$1607 & 1 	& 3201060101 	& 2020-03-13T04:47:22.249 	& 17.981 	& $0.80\pm0.19$ 	& $1.20$ 	& $12.3\substack{+2.0\\-1.9}$ 	& $-7.85\substack{+0.13\\-0.14}$ 	& $0.80$ 	& $-7.67\substack{+0.13\\-0.14}$ \\ 
& ...	& ... 	& ... 	& ... 	& ... 	& ...  & ... 	& ... & ... & ... \\ 
& 27 	& 3556015201 	& 2020-08-25T09:48:03.712 	& 53.695 	& $0.46\pm0.27$ 	& $1.20$ 	& $7.3\substack{+1.1\\-1.1}$ 	& $-8.30\substack{+0.12\\-0.14}$ 	& $0.80$ 	& $-8.13\substack{+0.12\\-0.14}$ \\ 
\hline
Swift J1822.3$-$1606 & 1 	& 1020460110 	& 2018-04-11T19:53:02.682 	& 206.345 	& $0.21\pm0.17$ 	& $-$ 	& $-$ 	& $-8.43\substack{+0.13\\-0.16}$ 	& $-0.76\substack{+0.36\\-0.38}$ 	& $-8.43\substack{+0.12\\-0.13}$ \\ 
\hline
SGR 1830$-$0645 & 1 	& 3201810102 	& 2020-10-11T13:13:22.627 	& 14.583 	& $-0.14\pm0.37$ 	& $1.20$ 	& $5.3\substack{+1.0\\-1.0}$ 	& $-8.42\substack{+0.15\\-0.17}$ 	& $0.80$ 	& $-8.25\substack{+0.15\\-0.17}$ \\ 
& ...	& ... 	& ... 	& ... 	& ... 	& ...  & ... 	& ... & ... & ... \\ 
& 128 	& 4201810109 	& 2021-03-22T05:28:50.886 	& 28.740 	& $-0.27\pm0.29$ 	& $1.20$ 	& $4.9\substack{+0.8\\-0.7}$ 	& $-8.49\substack{+0.12\\-0.14}$ 	& $0.80$ 	& $-8.32\substack{+0.12\\-0.14}$ \\ 
\hline
1E 1841$-$045 & 1 	& 7020500115 	& 2024-08-21T05:47:16.995 	& 275.708 	& $-0.30\pm0.20$ 	& $1.20$ 	& $4.4\substack{+0.5\\-0.5}$ 	& $-8.91\substack{+0.09\\-0.09}$ 	& $0.80$ 	& $-8.72\substack{+0.09\\-0.09}$ \\ 
& ...	& ... 	& ... 	& ... 	& ... 	& ...  & ... 	& ... & ... & ... \\ 
& 15 	& 7582010112 	& 2024-11-09T19:06:31.623 	& 82.960 	& $-0.58\pm0.19$ 	& $1.20$ 	& $6.1\substack{+0.7\\-0.7}$ 	& $-8.62\substack{+0.10\\-0.10}$ 	& $0.80$ 	& $-8.43\substack{+0.10\\-0.10}$ \\ 
\hline
SGR 1935+2154 & 1 	& 3020560101 	& 2020-04-28T00:42:04.992 	& 200.675 	& $-0.17\pm0.15$ 	& $2.09\substack{+0.80\\-0.44}$ 	& $2.5\substack{+0.8\\-0.7}$ 	& $-8.66\substack{+0.12\\-0.12}$ 	& $0.21\substack{+0.34\\-0.35}$ 	& $-8.57\substack{+0.10\\-0.10}$ \\ 
& ...	& ... 	& ... 	& ... 	& ... 	& ...  & ... 	& ... & ... & ... \\ 
& 867 	& 5576010116 	& 2022-11-15T11:48:59.834 	& 222.401 	& $-0.37\pm0.21$ 	& $1.20$ 	& $4.7\substack{+0.6\\-0.5}$ 	& $-8.97\substack{+0.10\\-0.10}$ 	& $0.80$ 	& $-8.78\substack{+0.10\\-0.10}$ \\ 
\hline
1E 2259+586 & 1 	& 7020600108 	& 2024-12-27T21:14:02.224 	& 236.260 	& $-1.00\pm0.00$ 	& $1.20$ 	& $1.8\substack{+0.2\\-0.2}$ 	& $-9.15\substack{+0.09\\-0.10}$ 	& $0.80$ 	& $-8.98\substack{+0.09\\-0.10}$ \\ 
\hline
PSR J1846$-$0258 & 1 	& 3033290103 	& 2020-08-05T17:08:55.807 	& 20.480 	& $0.12\pm0.25$ 	& $1.20$ 	& $13.4\substack{+1.7\\-1.6}$ 	& $-7.91\substack{+0.10\\-0.11}$ 	& $0.80$ 	& $-7.72\substack{+0.10\\-0.11}$ \\ 
& ...	& ... 	& ... 	& ... 	& ... 	& ...  & ... 	& ... & ... & ... \\ 
& 4 	& 3598010801 	& 2020-07-25T19:38:03.338 	& 11.285 	& $0.25\pm0.34$ 	& $1.20$ 	& $12.8\substack{+2.3\\-2.2}$ 	& $-7.95\substack{+0.14\\-0.16}$ 	& $0.80$ 	& $-7.76\substack{+0.14\\-0.16}$ \\ 
\hline
\multicolumn{9}{l}{$^{a}$ Values of $kT$ or $R$ that could not be constrained during spectral fitting are marked with “$-$”} \\
\multicolumn{9}{l}{$^{b}$ Uncertainties are not shown here for clarity but are provided in the full catalog table.} \\
\end{tabular}
\end{sidewaystable}




\end{document}